\documentclass[twocolumn,amsmath,amssymb]{revtex4}
\usepackage{graphicx}

\begin{document}

\title{Optical Manipulation of Light Scattering in Cold Atomic Rubidium}

\author{R.G. Olave\footnote{rolave@odu.edu}, A.L. Win, Kasie Kemp, S.J. Roof, S. Balik, and M.D. Havey}

\affiliation{Old Dominion University, Department of Physics, Norfolk, Virginia 23529 \\
}

\author{I.M. Sokolov${}^{1,2}$ and D.V. Kupriyanov${}^{1}$}

\affiliation{$^{1}$Department of Theoretical Physics, State Polytechnic
University, 195251, St.-Petersburg, Russia \\ \small $^{2}$Institute for
Analytical Instrumentation, RAS, 198103, St.-Petersburg, Russia  }

\begin{abstract}
A brief perspective on light scattering in dense and cold atomic rubidium is presented.   We particularly focus on the influence of auxiliary applied fields on the system response to a weak and nearly resonant probe field.  Auxiliary fields can strongly disturb light propagation; in addition to the steady state case, dynamically interesting effects appear clearly in both the time domain, and in the optical polarization dependence of the processes.  Following a general introduction, two examples of features found in such studies are presented.  These include nonlinear optical effects in (a) comparative studies of forward- and fluorescence-configuration scattering under combined excitation of a control and probe field, and (b) manipulation of the spatial structure of the optical response due to a light shifting strong applied field.
\end{abstract}

\maketitle

\section{Introduction}

The research presented in this paper forms one part of the broad interface between atomic and condensed matter physics.  A particular focus is an overview of quantum optics in high-density and cold atomic matter, in which disorder-driven electromagnetic interactions can develop strongly correlated character \cite{Kaiser}.  These emergent effects are typified by Anderson localization of light, the onset of random lasing and collective interatomic interactions such as super and sub radiance involving the system particles as a whole.   The overriding theme is then investigation of the physics of strongly correlated atomic-radiative systems, including the collective properties and phase transitions in ultracold atomic vapor. A closely related secondary theme appears when we consider the resonant interparticle interactions among the system constituents. Then the macroscopic optical response is modified from the microscopic single atom Lorentzian response to include so-called local field shifts, including the collective Lamb shift \cite{Friedberg1,Manassah1}.  It is important to note that much of the phenomenology appears most strongly in high density gases at low temperatures.  By this we mean that the temperatures on the order of 100 $\mu K$ or less, and atomic densities are characteristically greater than $10^{13}$ atoms/$cm^3$.  For such densities there are typically several atoms in a volume $(\lambda / 2 \pi)^3$, where $\lambda$ is the wavelength of light in the atomic medium.

We generally emphasize that the processes of emission and absorption of radiation in dense gases are quite representative of near-resonant many body interactions among the fields and the atoms in the ensemble \cite{Friedberg1,Manassah1}.  In the approximation where the atoms have a pairwise interaction of the dipole type, there contributes both a long range radiative interaction decreasing as 1/r with increasing atom-atom separation r, and short range interactions of the form of the Coulomb electrostatic dipole dipole interaction.  Among other effects which appear when virtual photons are exchanged among the different atoms in the system there emerges a collective Lamb shift which is essential to description of the global response of the system.  In general, the collective aspects yield a modification of the well-known local field shift \cite{Keaveney,Kampen} of the resonance frequency \cite{Juha}, and (with proper state preparation) an associated superradiant emission from the system \cite{Scully1,Rohlsberger1}.

In this paper we focus on the interaction of a weak and near resonance probe beam with an ensemble of cold atoms.  Even more specifically, we consider the influence of a second auxiliary field on the optical response of the probe.  Overall, we study influence of auxiliary fields in both the time domain and the dependence on the polarization of both detected light and the relative optical polarization of the auxiliary and probe fields.   There are two cases we consider.  In the first, the spectral structure of the atomic susceptibility is modified by the auxiliary field.  This is the case for the reported experimental results, and associated theoretical calculations, on polarized fluorescence excited in the so called lambda configuration characteristically employed in studies of electromagnetically induced transparency.  In the second, we consider modification of the spatial properties of an ultracold ensemble by an auxiliary field which produces a spatially varying light shift of the atomic levels.  This approach allows simulation of a quasi one dimensional spatial configuration of atoms which, for instance, can have applications to studies of superradiance \cite{Dicke}, Anderson localization \cite{Anderson} of light and random lasing \cite{Kaiser1,Cao,Letokhov,Gerasimov} in reduced spatial geometry.

\section{Illustrative Topics}

\subsection{Polarized fluorescence studies under conditions of electromagnetically induced transparency in cold atomic rubidium}

\subsubsection{Introduction}
In this section we report on a primarily experimental study directed towards optical control of weak
localization \cite{Sheng1,Akkerman1,Labeyrie2,Bidel,Kupriyanov1} in ultracold atomic vapor.  Generally, control is achieved by application of a dressing field to the atomic medium in the presence of the weaker probe incident radiation \cite{Lukin1,Lukin2}.  Such a configuration has a wide range of applications \cite{Lukin1,Chaneliere1,Matsukevich1,Darquie,Harris1,Braje1,Budker1,Budker2} in nonlinear optics \cite{Braje1,Braje2,HKang1,HKang2}, quantum information \cite{Lukin1,Chaneliere1,Matsukevich1}, single photon manipulation
\cite{Lukin1,Chaneliere1,Matsukevich1}, and precision magnetometry \cite{Budker1,Budker2}.  Here we consider manipulation of the weak probe beam and light scattered from it, thus undergoing multiple scattering in the weak localization regime. The particular scheme used is not optimum for large orders of multiple scattering, as the amount of multiple scattering is limited to at most several orders by the low optical depth on the relevant probe transitions, and by spontaneous emission into dark states which are only weakly coupled by either the control or probe field. In the experiments reported here we compare measured and theoretical polarization dependent forward and sideways scattered light. Finally we note that in these studies we do not use relative phase locked pump and control fields, thus realizing quite a bit shorter, and more readily measurable, time scales for the time evolution of both the probe and control field polarized fluorescence, and the forward scattered light.

In the remaining sections, we first describe the experimental approach.  This is followed by a presentation of our
investigation of the forward scattered probe light and the light polarization dependence of the scattered light in a ninety degree fluorescence geometry. Comparison and discussion is made through theoretical results \cite{Sokolov1,Datsyuk1,Datsyuk2} obtained for a model of the experimental configurations.

\subsubsection{Experimental Configuration}
The sample of cold $^{87}$Rb gas is obtained by cooling and loading the atomic vapor into a magneto optical trap (MOT)\cite{Metcalf}, which operates near the closed hyperfine transition $F = 2  \to  F^{\prime} = 3$. The trapping laser is derived from an external-cavity diode laser (ECDL) system in a master-slave configuration, and is split into three pairs of retro reflected beams, delivering an estimated trapping laser intensity of $ \simeq 40$ mW/cm$^2$.  To prevent optical pumping of the atomic vapor into the lower energy $F = 1$ hyperfine level, a repumper laser is set to the $F = 1 \to F^{\prime} = 2$
hyperfine transition. The trap and repumper lasers can be frequency tuned and switched on or off through acousto optical
modulators (AOM). Fluorescence imaging measurements show that the sample has an effective Gaussian atom distribution, which can be described by $\rho(r) = \rho_o e^{-r^2/2r_o^2}$; for this sample $r_o \approx 0.3$ mm.  The temperature of the $^{87}$Rb atomic gas was determined via ballistic expansion measurements, and is estimated to be $\sim 100\ \mu$K.  Finally, a maximum optical depth of $b \sim 10$ on the trapping transition was determined through absorption imaging measurements, resulting in a peak atomic density of $\rho_o \sim 9 \cdot 10^{10} $ atoms/cm$^3$.

In this experiment, several overlapping Zeeman $\Lambda$ schemes operating within a three energy-level configuration are shown in Fig. 1. There the resonance transition frequencies are given by $f_{11}$ for the $F = 1  \to  F^{\prime} = 1$ transition, and $f_{21}$ for the $F = 2  \to  F^{\prime} = 1$ transition.  Before any measurement, the repumper laser is turned off for 1 ms, and in this period approximately 86\% of the atoms are optically
pumped to the $F = 1$ hyperfine ground state.  Initially, an intense linearly
polarized control laser beam with Rabi frequency $\Omega_c$ couples on
bare atomic resonance the $F = 2$ ground state to the $F^{\prime} = 1$ excited state
($\Delta_c = 0$, where $\Delta_c = f_c - f_{21}$ is the control field detuning from resonance). The medium
is then probed by a weak pulse with orthogonal linear polarization and Rabi
frequency $\Omega_p$, with a frequency detuning from the bare atomic resonance given
by $\Delta_p$.   Here $\Delta_p = f_p - f_{11}$.

\begin{figure}
{\includegraphics{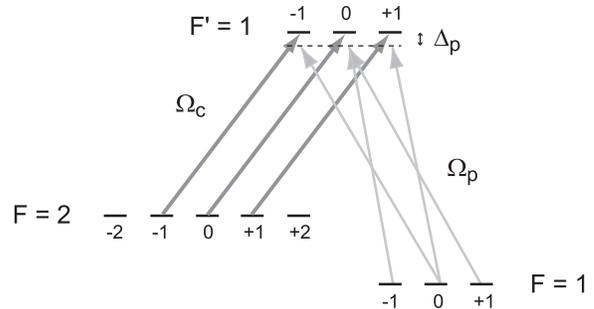}}
\caption{Relevant hyperfine lambda scheme.  The near-resonant strong
coupling laser has Rabi frequency $\Omega_c$.  The probe laser has
Rabi frequency $\Omega_p$, and detuning from the bare atomic resonance
of $\Delta_p$.}
\label{Figure1}
\end{figure}

\begin{figure}
{\includegraphics{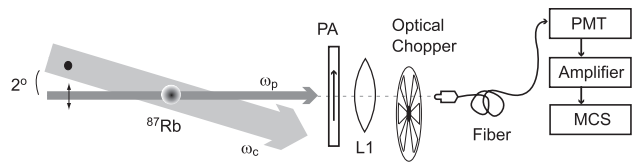}}
\caption{Experimental setup for detection of transmission of a
probe pulse through the atomic gas. The coupling laser
has a frequency $f_c$, while the probe laser has a frequency
$f_p$.  A lens L1 collects the probe pulse transmitted
through the Rb cloud and launches the beam into a fiber.  PA is a
polarization analyzer, PMT is a photomultiplier tube, and MCS is a
multichannel scaler.}
\label{Figure2}
\end{figure}

The control and probe lasers come from separate ECDL systems,
with a combined root-mean-square bandwidth of 1.3 MHz, determined via optical
heterodyning measurements.  The control beam has a
Gaussian spatial intensity profile, with a radius of $r_{\omega c}
= 0.9$ mm, and is closely uniform in intensity over the volume of the MOT.
The probe beam also has a Gaussian spatial intensity
profile, and a radius of  $r_{\omega p} = 0.2$ mm. The probe is directed through the central region of the atomic sample.

\subsubsection{Results and Discussion}
In this section we first present our results for forward scattering geometry, where we study the time-dependent optical behavior of a near resonance probe pulse transmitted through a $^{87}$Rb vapor cloud dressed by an optical control field.  This is followed by time-resolved observation of the polarized probe and control field induced sample fluorescence in a $90^\circ$ geometry.

\subsubsection{Forward Scattering Geometry}

A schematic setup for the detection of the transmitted probe pulse
is shown in Fig. 2. A coupling laser
with frequency $f_c$ together with a probe laser of frequency $f_p$ forms a coupled set of near-resonance lambda configurations, and the forward scattered probe light is detected. To minimize background and to avoid saturation of the detector, the coupling laser
has been offset by about 2$^\circ$ from the probe pulse propagation
axis.  A polarization analyzer (PA) selects light with the same
polarization as the incident probe pulse, thus strongly suppressing the control field background at the detector. A lens L1 focuses the transmitted light into a multimode fiber which further directs the light to an infrared-sensitive
photomultiplier tube (PMT).  The time-resolved signals are then amplified using a fast preamplifier and recorded by a multichannel scaler.  The system time resolution is about 5 ns.

\begin{figure}
{\includegraphics{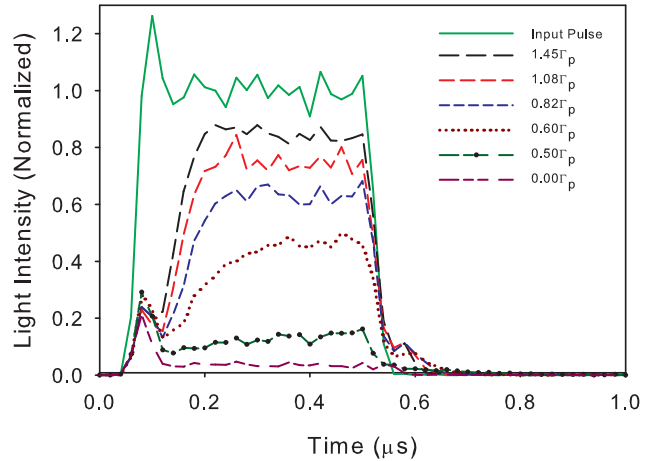}}
\caption{Measurement of the time evolution of the transmitted probe
pulse for varying $\Omega_c$. The nearly rectangular probe pulse
is 500 ns long.  Data is taken at $\Delta_p = \Delta_c = 0$, for
an atomic sample with optical depth of $b \simeq 3$.   The slow
light components can be observed as well. The probe pulse in the absence of a sample,
and the probe pulse when there is no coupling laser ($\Omega_c =
0.0 \ \Gamma_p$) are included for comparison. Here, $\Gamma_p = 3.8 \cdot 10^{7} s^{-1}$.}
\label{Figure3}
\end{figure}

\begin{figure}
{\includegraphics{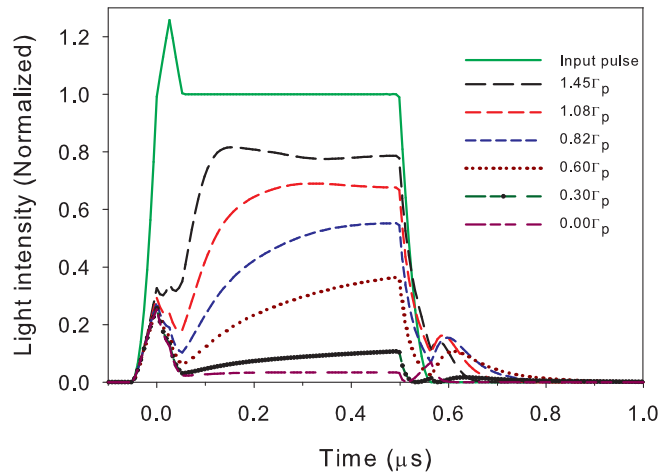}}
\caption{Theoretical calculations of the time evolution of the
transmitted probe pulse for varying $\Omega_c$, as indicated in the caption. The probe pulse is approximately
500 ns long, at $\Delta_p = \Delta_c = 0$, for an atomic sample
with optical depth of $b = 3.4$.  The calculations assume a ground
states decoherence rate of 0.07 $\Gamma_p$.  Here, $\Gamma_p = 3.8 \cdot 10^{7} s^{-1}$.}
\label{Figure4}
\end{figure}

For a resonant optical depth of $b \simeq 3$ on the probe transition, measurements were taken to compare the
time evolution and intensity of the transparency of a 500 ns long probe pulse
as a function of the control field intensity $\Omega_c$; the experimental
results, for a resonant probe $\Delta_p = \Delta_c = 0$, can be seen in Fig.
3. We point out that, strictly speaking this detuning relation does
not precisely define the point of maximal transparency because of the
additional light shift induced by the coupling with other upper-state hyperfine
sublevels \cite{Giner}. However in the discussed experimental
conditions, when the Rabi frequency has the same order as the natural decay rate
$\Gamma_p = 3.8 \cdot 10^{7} s^{-1}$ this correction is very small. The measurements are time averaged so as to have a
20 ns time resolution; this improves the signal to noise ratio, while at the same time giving a temporal resolution shorter than the natural lifetime of 26 ns. Clearly, the transparency develops for all employed values
the coupling field strength, but it reaches a steady state faster for greater
values of $\Omega_c$. In these data the delayed pulses associated with slow
light can be seen as well. It should be
noted that the incident probe pulse is not exactly temporally square; it has an
estimated rise and fall time of 40 ns.  The pulse initially overshoots (about
10\% above the main signal), this is due in part to initial overshooting of the
acousto-optic modulator (AOM) that controls the switching of the pulse. This
overshoot does not take place when the AOM switches off.

Theoretical calculations for the transmission of the probe pulse under similar conditions are presented in Fig. 4.  The calculational approach we have used is presented in detail in several of our earlier papers \cite{Sokolov1,Datsyuk1,Datsyuk2}, and is not reiterated here.  We do point out that the probe pulse shape, including the mentioned overshoot, and the measured rise and fall times, is built into the calculation.   In comparing Figs. 3 and 4, we see overall qualitative agreement between the experimental and theoretical time responses at the various control field Rabi frequencies.

\begin{figure}
{\includegraphics{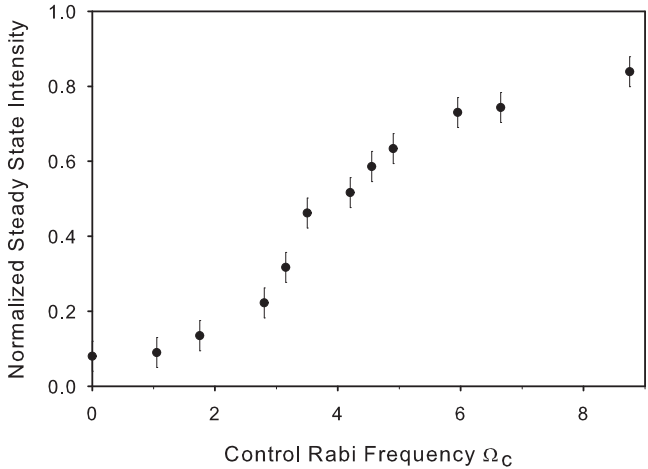}}
\caption{Intensity of the steady-state transmitted probe signal  as
a function of $\Omega_c$.  The probe and control fields are on
resonance ($\Delta_p = \Delta_c = 0$).  The signals are normalized
to the intensity of the undisturbed probe pulse.}
\label{Figure5}
\end{figure}

\begin{figure}
{\includegraphics{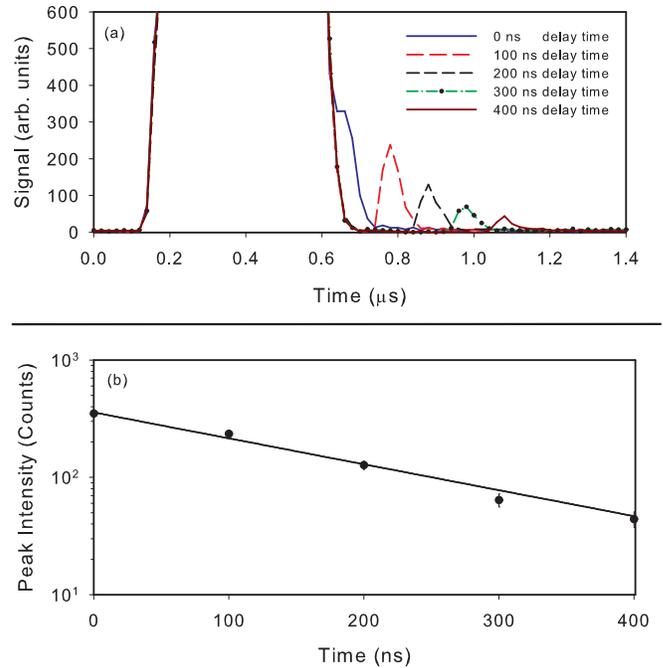}}
\caption{Probe pulse light storage and retrieval with variable control
field delays for $\Omega_c = 1.45 \ \Gamma_p$.  (a) The control
field is switched off at $t = 0.6\ \mu s$, and turned on after
different time intervals.  (b) Comparison of the retrieved probe
signal for variable storage times.  A fit to the data yields
a 1/$e$ decay rate of the atomic coherence of 195 ns. Here, $\Gamma_p = 3.8 \cdot 10^{7} s^{-1}$.}
\label{Figure6}
\end{figure}

From results similar to those of Fig. 3, the steady state response of the configuration as a function of the control field Rabi frequency may be extracted. Fig. 5 shows such results for Rabi frequencies $\Omega_c$ ranging from well smaller than the natural width up to about 8 MHz.   For this figure, the signals have been normalized such that the steady-state intensity of the incident probe pulse corresponds to 1 (100\% transparency).  The uncertainty in the data comes primarily from signal counting statistics and from run-to-run variations in experimental conditions.

Figure 6 (a) presents a set of measurements taken,  for a
control field Rabi frequency $\Omega_c = 1.45 \ \Gamma_p$, to investigate storage and retrieval of light pulses.  For these
measurements the control and probe fields are simultaneously turned off (at $t
= 0.6\ \mu s$), and the control field is switched back on after some variable
time delay.  As can be seen, when the control field is turned back on, some portion of the
stored light pulse is retrieved.  Figure 6 (b)\ is a plot of the integrated transmission of the probe
pulse versus storage time.  A fit to the data yields a 1/$e$ decay time of the
atomic coherence of 196 ns, or a decoherence rate on the order of 0.9 MHz.  In order to physically understand this relatively large decoherence rate, we consider that when the pump and control fields are turned off, a dark state polariton is formed with some degree of efficiency.  Once the probe and control fields are off, the quasiparticle decays relatively slowly in the dark with decoherence time limited mainly by residual magnetic fields over the volume of the atomic sample.  However, when the control field is turned back on, in order to regenerate the probe optical field, the control field phase has evolved from when it was turned off. For each realization of the experiment, this phase in relation to that of the created polariton will be different.   This phase evolution leads to a decay of the coherence that is proportional to the linewidth of the control laser field ($\sim$ 1 MHz), which becomes relevant when it is switched back on to regenerate the forward scattered light.

\subsubsection{Fluorescence Detection Geometry}

A schematic setup for the detection of fluorescence is shown in
Fig. 7. In these measurements, the linearly polarized control laser
with frequency $f_c$ is initially switched on,  and 50 $\mu$s later a weak laser probe pulse with frequency $f_p$
and orthogonal linear polarization probes the system.  A
linear polarization analyzer PA selects the polarization state of the light to be detected.
A lens L1 collects and collimates
fluorescence coming from the location of the MOT, and a lens L2
focuses the light into a multimode optical fiber which transfers the light to the PMT.  The
signal is then amplified and time-resolved using a multichannel
scaler.  Note that for these measurements, in contrast with the transmission experiments, there is no 2$^\circ$
offset between the control and the probe beams.  The relative response of the optical and electronic detection system is calibrated for two orthogonal linear polarizations, corresponding to directions collinear with, or orthogonal to, the probe or control field.  This allows meaningful measurements of the linear polarization degree of the time-resolved fluorescence to be made.

\begin{figure}
{\includegraphics{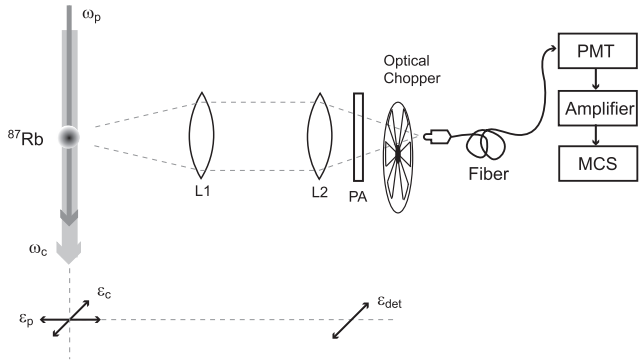}}
\caption{Experimental setup for detection of fluorescence from a
probe pulse propagating through the atomic gas.  The
coupling laser has a frequency $\omega_c$.  The probe laser has a
frequency $\omega_p$.  A lens L1 collects fluorescence emitted
from the atomic cloud, while a lens L2 focuses this light into a
fiber.  PA is a polarization analyzer, PMT is a photomultiplier
tube, MCS is a multichannel scaler.  The probe and control lasers
have linear and mutually orthogonal polarization.  The detected
polarization is variable, and is shown in the figure to be collinear with the control field polarization.
$\epsilon_p, \epsilon_c$, and $\epsilon_{det}$ are the probe,
control, and detected light polarization vectors respectively. Drawing not to scale. }
\label{Figure7}
\end{figure}

The fluorescence generated by optical excitation of the system by the probe and control fields, as a function of
the control laser Rabi frequency $\Omega_c$, is presented in Fig.
8.  In these measurements the control field is linearly polarized and perpendicular to the detector plane.  The detector linear polarization analyzer is set to maximally pass light polarized in the same direction as the control field.  For a control field with a low intensity ($\Omega_c <
\Gamma_p$), a slower build up of fluorescence is initially
observed, but the transparency is small and it has a narrow
frequency width, so the main contribution to the fluorescence
signal comes from scattered light having frequency components outside the transparency window.  On the other hand, for large
coupling field intensities ($\Omega_c > \Gamma_p$) the medium develops a high
degree of transparency for a wider frequency range, which
results in a suppression of fluorescence clearly observed in the data.

\begin{figure}
{\includegraphics{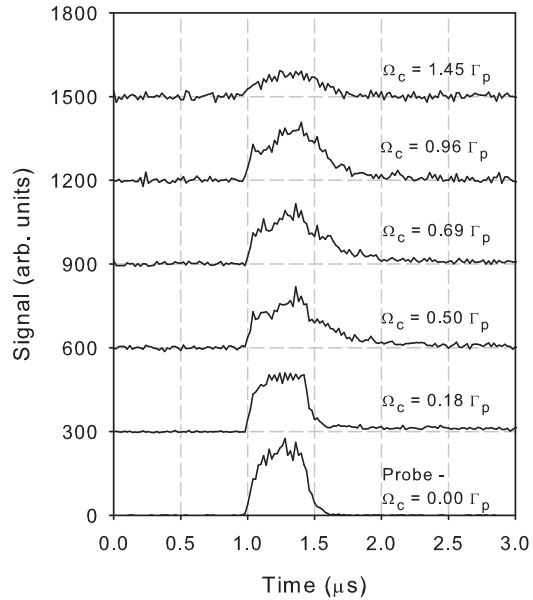}}
\caption{Probe fluorescence from an optically dressed $^{87}$Rb
cloud, as a function of the control laser Rabi frequency
$\Omega_c$.  The probe pulse is 500 ns long, on resonance
($\Delta_p = \Delta_c = 0$).  When $\Omega_c = 0$ a characteristic probe-only
fluorescence signal is detected, shown here for comparison.  The
control field is turned on 50 $\mu$s before the probe pulse, and
stays on for 60 $\mu$s. }
\label{Figure8}
\end{figure}

\begin{figure}
{\includegraphics{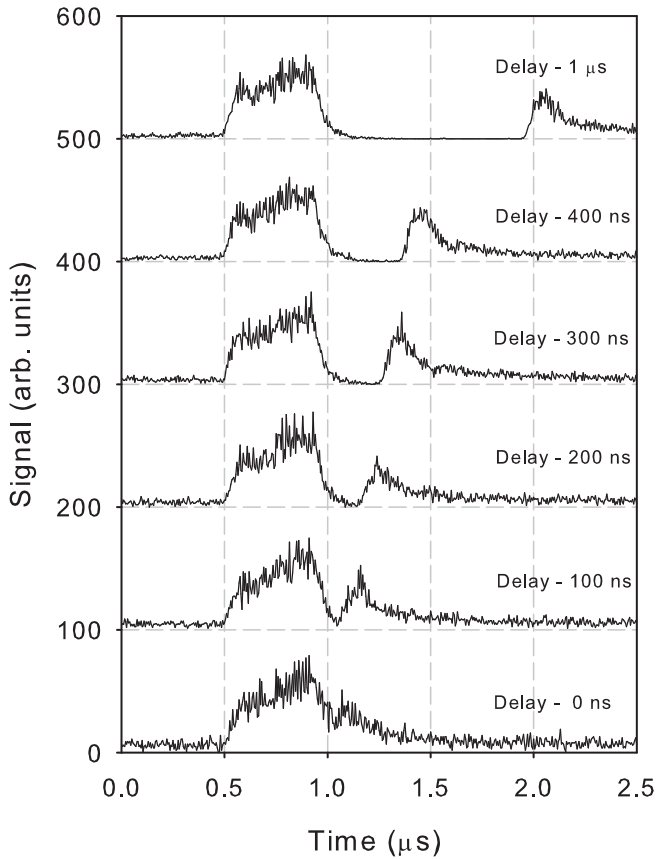}}
\caption{Probe fluorescence as the control field, with a Rabi
frequency of $\Omega_c = 0.69 \ \Gamma_p$, is turned off/on for
various time delays.  The probe pulse frequency is on resonance
($\Delta_p = 0$).  The slowly decaying signal obtained when the
control field is switched back on could be the result of optical
pumping of atoms into the $F = 1$ level by the control field.  Here, $\Gamma_p = 3.8 \cdot 10^{7} s^{-1}$. }
\label{Figure9}
\end{figure}

For weak control field Rabi frequencies, there is a slowly decaying portion of the fluorescence signal, as revealed by Fig.
9.  To explore this slow decay, we employed a protocol like that used in storage and retrieval type of measurements of slow fluorescence light, and similar to our transmission measurements (Fig. \ref{figure6}).  The results, for a control field Rabi frequency of $\Omega_c = 0.69 \ \Gamma_p$, with 5 ns resolution, are presented in Fig. 9.  In these
measurements, the control field was turned off at $t = 0.5\ \mu s$, and switched back on with variable time delays. We see that the obtained fluorescence signals, upon switching back on the control field, with no apparent amplitude decay, for periods of time much longer than the decay time associated with the coherence time in the vapor. As we saw earlier in this section, the 1/$e$ decay rate for the transmission measurements is about 200 ns.  Although there is a qualitative change in the shape of the generated fluorescence pulse on the time scale of 200 ns, the persistence of the main features of this signal, which may be generated for delay times greater than 10 $\mu s$, has a different origin.  We attribute these fluorescence signals to optical
pumping by the control field (initiated by spontaneous Raman scattering from the F = 1 level), which populates the $F = 2,\ m = \pm 2$ Zeeman levels (see Fig. \ref{figure1}).  The control beam has $\sim$ 0.1 \% of light linearly polarized orthogonal to the main beam, which is enough to slowly optically pump the atoms into the $F = 2$ level, and to also contribute to the fluorescence signal in the process. For the highest control field Rabi frequencies, optical pumping occurs more quickly, and this parasitic signal, although the mechanism remains active, is not apparent in the data taken here.  Such effects may occur in other circumstances, and can occur even under the circumstances when there is a high degree of phase coherence between the control field and the probe field.

We now turn to measurements of the polarization dependence of the time-resolved fluorescence, using the geometry of Fig. 7.  For comparison purposes, we point out that we have previously reported linear polarization measurements for the case of a weak probe beam only \cite{Balik1}. In these measurements, as in the others reported in this paper, the probe and control field have mutually orthogonal linear polarization directions.  In the fluorescence detection arm there is a linear polarization analyzer which may be rotated so as to change the detected state of polarization.   The detection plane, formed by the wave vectors of the probe and control fields, and the detected fluorescence direction, is horizontal.  Four different measurements are made.  In two of them, the probe field is vertical, the control field polarization vector is in the detection plane, and measurements are made of the fluorescence signals with the detector polarization analyzer either collinear with the probe field direction or perpendicular to it.  In the other two measurements, the control field polarization is vertical, and the probe field vector lies in the detection plane.   Again, measurements are made of the fluorescence signals alternately with the detector polarization analyzer collinear with the control field direction and perpendicular to it.  These polarization configurations are illustrated by inserts to Figs. 10 - 13.

\begin{figure}
{\includegraphics{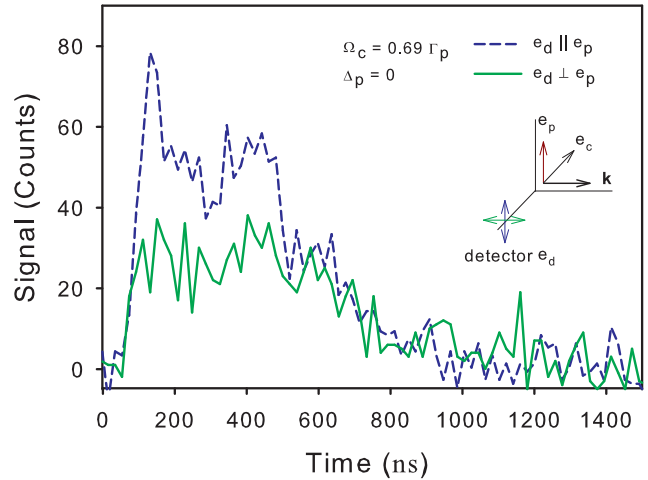}}
\caption{Experimental results for the time-dependent polarized fluorescence under similar conditions.  For these results the probe linear polarization is vertical while the control field linear polarization is horizontal. As indicated, two results are shown; in one case the linear polarization of the detected light is in the same direction as the probe linear polarization, while in the other they are orthogonal.}
\label{Figure10}
\end{figure}

\begin{figure}
{\includegraphics{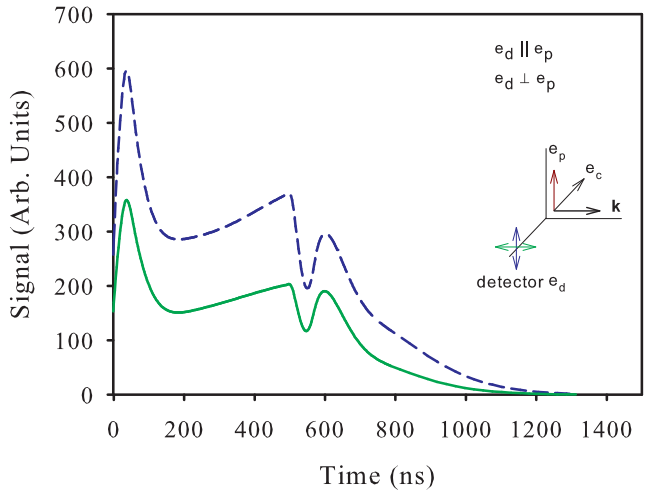}}
\caption{Theoretical calculations of the time-dependent polarized fluorescence under similar conditions.  For these results the probe linear polarization is vertical while the control field linear polarization is horizontal. As indicated, two results are shown; in one case the linear polarization of the detected light is in the same direction as the probe linear polarization, while in the other they are orthogonal.}
\label{Figure11}
\end{figure}

The measured time-resolved fluorescence signals with the $\emph{probe field}$ vertically polarized, and for two orthogonal states of detected light linear polarization, are shown in Fig. 10.  In these measurements the data, taken in 5 ns bins, are regrouped into 20 ns bins in order to improve the signal to noise ratio.   In the figure, the overall evolution of the intensity is a rise (with an overshoot, as discussed previously) to an approximate steady state followed by decay of the signal after the probe is extinguished.   We also see that the fluorescence has a nonzero degree of linear polarization, with an average linear polarization degree in the range $30 - 35 \%$, while the probe and control field are both on. Here, as is customary, the linear polarization degree is defined as the difference in the measured intensities for the two orthogonal polarization states, normalized to the sum of these intensities.  We first note that if this excitation were purely from the probe $F = 1 \rightarrow F' = 1 \rightarrow F = 1$ transition, then the linear polarization degree would be about $33 \%$, well within the uncertainty of the measured polarization as in Fig. 10.  We point out that if we include contribution from the spectrally unresolved $F = 1 \rightarrow F' = 1 \rightarrow F = 2$ Raman transition, the linear polarization would be reduced to 3/11 ($\sim 27$ \%).  In Fig. 11 we present theoretical calculations of the time resolved fluorescence.  The results are calculated for the same control field Rabi frequency as the experimental data; the relative control and probe beam sizes are also matched to the experimental results.  We see in the theoretical results good correspondence with respect to the measurements when the probe field remains on.  However, once the field is extinguished the experimental results show that the remaining transient fluorescence signals are unpolarized, within the experimental uncertainty, whereas there remains a clear polarized component in the theoretical results.  We can understand these results by considering that the fluorescence signals must arise from components of the probe field that do not spectrally overlap the transparency window.   With reference to Fig. 5, for the selected control field Rabi frequency of $\Omega_c = 0.69 \ \Gamma_p$, a significant fraction of the probe field satisfies this condition.  This signal would be expected to have a linear polarization degree close to the estimates provided above for single atom scattering.

\begin{figure}
{\includegraphics{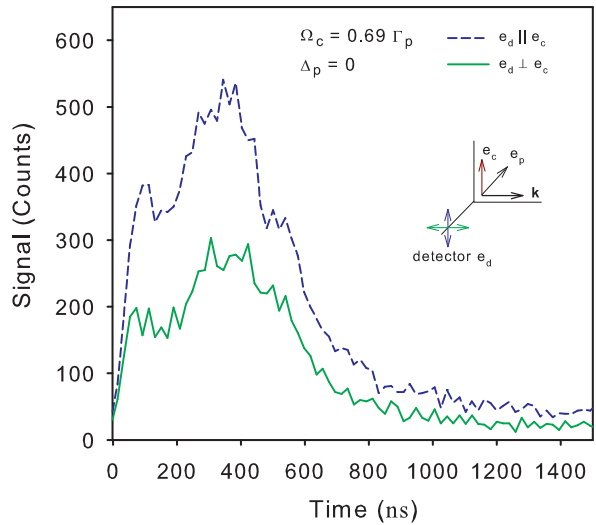}}
\caption{Experimental results for the time-dependent polarized fluorescence under similar conditions.  For these results the control field linear polarization is vertical while the probe field linear polarization is horizontal. As indicated, two results are shown; in one case the linear polarization of the detected light is in the same direction as the control field linear polarization, while in the other they are orthogonal.}
\label{Figure12}
\end{figure}

\begin{figure}
{\includegraphics{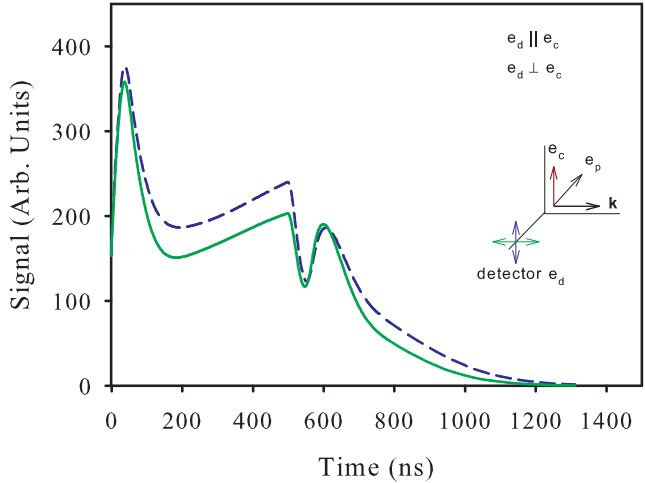}}
\caption{Theoretical calculations of the time-dependent polarized fluorescence under similar conditions.  For these results the control field linear polarization is vertical while the probe field linear polarization is horizontal. As indicated, two results are shown; in one case the linear polarization of the detected light is in the same direction as the control field linear polarization, while in the other they are orthogonal.}
\label{Figure13}
\end{figure}

The measured time-resolved fluorescence signals with the $\emph{control field}$ vertically polarized, and for two orthogonal states of detected light linear polarization, are shown in Fig. 12.  As for the data in Fig. 10, the time resolution is limited by the bin size of 20 ns.  In the figure, the overall evolution of the experimental intensity is more structured than in the previous case.   We also see that the fluorescence is linearly polarized over the entire time evolution of the signals, even after the probe and control fields are shut off; for the whole span of the data, the average linear polarization degree is about $35 \%$. Theoretical results are calculated for the same control field Rabi frequency as the experimental data; the relative control and probe beam sizes are also matched to the experimental results.  As in Fig. 11, we see in Figure 13 that the theoretical results have a qualitative correspondence with respect to the measurements.  In contrast to the results of Fig. 10 and Fig. 11, we observe in Fig. 12-13 that the experimental and theoretical polarization of the fluorescence signals are quite closely matched, even when the probe and control fields are turned off.

\subsubsection{Conclusion}
In this section, we have discussed several aspects of near resonance lambda scheme optical excitation of a cold sample of $^{87}$Rb atoms.  Forward scattering measurements are used to characterize the optical response of the dressed atomic sample.  Further experiments in a right angle fluorescence geometry explore the linear polarization degree of scattered light for two different spatial configurations of the pump and control fields with respect to the fluorescence detector.  These measurements show a significant linear polarization degree that is essentially constant while the probe and control fields are both incident on the atomic sample.  Once the fields are extinguished the initial polarization degree of the polarized probe fluorescence rapidly decays to zero, while that of the polarized control field fluorescence is maintained.  These dynamical polarization effects are in reasonable agreement with theoretical modeling of the processes.

\subsection{A High-Density Optical Atomic Trap for Study of Cooperative Interactions in Light Scattering}
\subsubsection{Brief Introduction}

In this portion of the paper, we give an overview of how we produce a quasi one dimensional configuration of high density and cold $^{87}$Rb. Our approach builds on earlier work associated with a $CO_2$ laser based quasielectrostatic trap \cite{Balik,Balik2}.  This modified trap may be used for a number of experiments including measurement of the forward coherent and incoherently scattered light, and experiments directed towards Anderson Localization of light in quasi one dimension.  Here we focus on how we characterize the atomic sample, and illustrate with one method how a quasi one dimensional geometry can be created.

\subsubsection{The Magneto Optical Trap}
We begin with atoms collected and cooled in a magneto optical trap (MOT) well described by a density distribution $n(r)$ having a peak density $n_0$ and radius $r_0$ given by

\begin{equation}
\label{density}
n(r)=n_0\exp\left({-\frac{r^2}{2r_0^2}}\right).
\end{equation}

In order to measure the number of atoms in the MOT, we determine the absorption of light from a near-resonance probe beam with intensity $I_0$ that is incident on the sample.  From the Beer-Lambert Law, the transmitted light through the sample is

\begin{equation}
I_T=I_0 e^{-b},
\end{equation}

with $b$ being the optical depth.  For a not-too-optically deep sample

\begin{equation}
b(r,\delta)=\sqrt{2 \pi} r_0 n(r) \sigma(\delta)
\label{OD}
\end{equation}

where
\begin{equation}
\label{sigma}
\sigma(\delta)=\frac{\sigma_0}{1+\left(\frac{2\delta}{\gamma}\right)^2}
\end{equation}

is the scattering cross section with a laser detuning $\delta$ from resonance; $\gamma$ is the lifetime of the excited state. The expression for the cross section applies when the density is not too large, and when interatomic interactions may be ignored.  Initially, a charge coupled device (CCD) camera takes an image of the transverse spatial probe intensity distribution. Then, an image is taken with the atomic sample present, yielding an absorption image. After background correction, the natural log of the ratio of these two images yields

\begin{equation}
\ln \frac{I_T(r,\delta)}{I_0}=-b(r,\delta).
\end{equation}

This image is well fit to a Gaussian curve, with the amplitude equal to the peak optical depth and the $1/e^\frac{1}{2}$ radius giving $r_0$ of the sample. Integrating Equation \eqref{density}, we find that the total number of atoms in the MOT is

\begin{equation}
N_{MOT}=\left(2\pi\right)^{3/2}n_0\;r_0^3
\label{number}
\end{equation}

Using an on-resonance probe $(\delta=0)$, substitution of Eqs. \eqref{OD} and \eqref{sigma} into Eq. \eqref{number} yields

\begin{equation}
N=2\pi \frac{r_0^2\; b_0}{\sigma_0}
\end{equation}

\begin{equation}
\sigma_0=\frac{2F'+1}{2F+1}\frac{\lambda^2}{2\pi}
\end{equation}

Probing the $5^2S_{1/2}, F=2 \rightarrow 5^2P_{3/2}, F'=3$ transition with a laser of wavelength of 780 nm, we measured that the MOT typically has $\sim 5 \cdot 10^7$ atoms.

\subsubsection{High Density Far Off Resonance Trap; Loading and Characterization}
In order to reach the regime of higher atomic density, the atoms collected in the MOT are partially transferred to a far off resonance trap (FORT).  This trap is formed at the focus of a fiber laser beam operating around 1.06 $\mu m$ and having a power of several watts.  To obtain optimal loading, good spatial overlap between the FORT beam and MOT atoms must be assured. This is done by controlling a wide parameter space that includes repumper beam power and turn-off time, MOT beam detuning, FORT power and alignment, and loading time.  The most important of these is repumper power during loading.  It is extremely sensitive to the optimal setting due to the need to reduce radiation pressure within the MOT and the ability to pump the atoms down into the F=1 ground state, which has a small inelastic collisional cross-section.  Typically, the repumper power is reduced to $\sim$30 $\mu$W and has a turn-off time that precedes that of the MOT beam by 3 ms to help with the pumping process to the F=1 state.  The MOT beam is detuned to also help with the compression of the MOT atoms around the FORT beam and a detuning of $\sim$10$\gamma$ is generally optimum.  The fiber laser that provides the FORT potential, which can be increased up to 35 W, is operated around 2 W.  This power setting is sufficient as the beam is focused to $\sim$18 $\mu$m giving a trap depth of 670 $\mu$K.  As referenced in \cite{Kuppens}, the FORT beam focus should be offset from the center of the MOT atoms as the best loading is achieved in the throat region of the trap.  This entire experimental protocol lasts for 70 ms, typical of most FORTs \cite{Minarni}, at which point the MOT beams are shut off along with the magnetic field and a fraction of the MOT atoms thermalize by collisions in the dipole potential trap.

By comparing the fluorescence of the MOT and FORT under optically thin conditions, we can then find the transfer efficiency and number of atoms successfully shifted from the MOT to the FORT. With a typical transfer efficiency of $\sim 5\%$ the FORT has $\sim 2.5 \cdot 10^6$ atoms. The number of atoms in the FORT decreases with longer hold times, as seen in Fig. 14. This is due to collisions between atoms in the FORT and background gas atoms in the vacuum chamber.

After finding the number of atoms in the FORT, it is necessary to know the temperature of the atom sample. The FORT has a bi-Gaussian distribution of atoms

\begin{equation}
n(r,z)=n_0\exp\left(-\frac{r^2}{2r_0^2}-\frac{z^2}{2z_0^2}\right)
\end{equation}

and, once thermalized, a Maxwell-Boltzmann distribution of velocities \cite{Metcalf}

\begin{equation}
f(v) d^3v=\left(\frac{2\pi k_B T}{m}\right)^{-3/2} \exp \left(- \frac{m v^2}{2 k_B T} \right) d^3v.
\end{equation}

Here, $m$ is the mass of an individual atom, $k_B$ is the Boltzmann constant, and $T$ is the average temperature of the sample.  When atoms in the FORT are released from the trap, their position will change with time according to

\begin{equation}
\vec{r}=\vec{r}\,'+\vec{v}\;t.
\end{equation}

The time-dependent distribution of atoms, which is itself a bi-Gaussian distribution, is given to a good approximation by

\begin{equation}
n(\vec{r},t)=\iiint\limits_{-\infty}^{\hspace{10pt}\infty}d^3v\;f(v)\; n(\vec{r}-\vec{v}\;t).
\end{equation}

If we allow the cloud of atoms to ballistically expand, we can find the two characteristic radii of the sample as a function of time. The temperature is then proportional to the rate of change of the radii.

\begin{equation}
\begin{array}{c}
r^2=r_0^2+\frac{k_B T}{m}t^2 \\
z^2=z_0^2+\frac{k_B T}{m}t^2
\end{array}
\end{equation}

The typical temperature of the FORT is $\sim 100\; \mu K$ and that of our MOT is $\sim 25\; \mu K$.
	
Finally, to find the initial radii of the FORT potential, and thus the effective volume, we utilize parametric resonance.  Because of the low atom temperature in comparison with the trap depth, the distribution of atoms in the trap is concentrated near the bottom of the confining dipole potential.  Then the radial and axial radii for the sample are significantly smaller than the beam waist and Rayleigh range, and the potential can be approximated as a harmonic oscillator

\begin{equation}
U(r,z)\approx -U_0 \left(1-\left(\frac{z}{z_r^2}\right)^2-2\left(\frac{r}{w_0^2}\right)^2\right).
\end{equation}

This can be rewritten to include a radial oscillation frequency $\omega_r$ and transverse frequency $\omega_z$

\begin{equation}
U(r,z)=-U_0+\frac{1}{2}m\;\omega_r^2\;r^2+\frac{1}{2}m\;\omega_z^2\;z^2
\end{equation}

with harmonic oscillation frequencies of

\begin{equation}
\begin{array}{c}
\omega_r=\sqrt{\frac{4U_0}{m\omega_0^2}}\\
\omega_z=\sqrt{\frac{2U_0}{mz_r^2}}.
\end{array}
\end{equation}

We begin with a trapping beam with an intensity described by a Gaussian distribution, and then apply a small sinusoidal modulation to this intensity for a fixed time while the atoms are held in the potential. The strength of the modulation is $h$, and the frequency is $\omega$.

\begin{equation}
I(r,z,t)=I(r,z)+I_0\;h\cos \left( \omega \; t \right).
\end{equation}

The one dimensional Mathieu equation describes the motion of atoms within the trap

\begin{equation}
\ddot{r}+\kappa\;\dot{r}+\omega_r^2\;r\left(1+h\;\cos\left(\omega\;t\right)\right)=0
\end{equation}

with a relaxation rate $\kappa$.  A resonance occurs when

\begin{equation}
\omega=2\omega_r/n, \qquad n=1,2,\dots.
\end{equation}

We can detect when we are on resonance by recording the intensity of the fluorescence as a function of frequency.  The strongest response occurs at $\omega=2\omega_r$. For our FORT, the radial frequency is on the order of $2\pi\cdot 4$ kHz and the transverse frequency is $2\pi\cdot 45$ Hz.  With these frequencies and the temperature determined from ballistic expansion, we can calculate the radial and transverse radii of the atomic sample.

\begin{equation}
r_0=\sqrt{\frac{k_b\;T}{m\omega_r^2}}; \qquad z_0=\sqrt{2}\pi\;r_0\;\omega_0/\lambda
\end{equation}

While the trap has a beam waist of $\sim 17.6\; \mu$m and Rayleigh range of $\sim 0.92$ mm, the FORT atom distribution has a radial Gaussian radius of $\sim 3.4\; \mu$m and a transverse Gaussian radius of $\sim 250\; \mu$m.  Thus, the peak density of our sample of atoms is $6 \cdot 10^{13}$ atoms/cm$^3$.

\begin{figure}
{\includegraphics{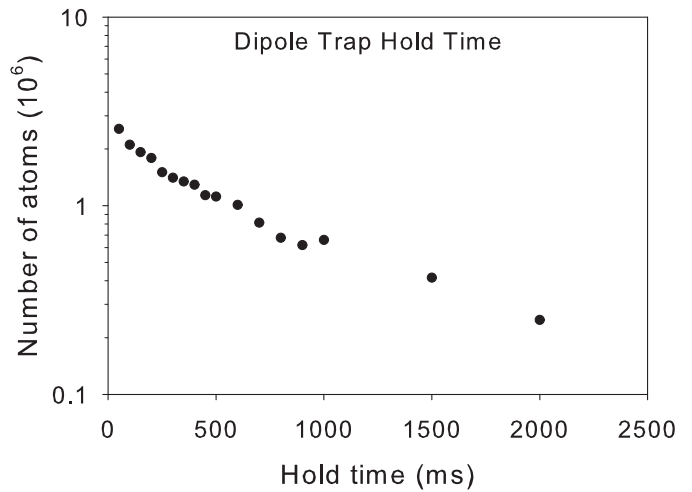}}
\caption{Number of atoms remaining in the dipole trap as a function of time.   The atom number decreases due to collisional ejection of the atoms from the trap, mainly by hot Rb atoms in the chamber. }
\label{Figure14}
\end{figure}

\subsubsection{Creating and Probing a Dense and Cold Quasi-One-Dimensional Ensemble}
Cold atomic samples with high densities have numerous valuable and interesting properties. Our immediate interests are in three areas; (a) spectral and density dependence of the forward scattered light intensity, (b) demonstration and characterization of Anderson localization of light in a quasi one dimensional (Q1D) geometry, and (c) study of random lasing in Q1D.  These experiments require creation of a Q1D channel through the atomic sample.   One way to accomplish this is described below, and an image of such a realization is shown in Fig. 15.

\begin{figure}
{\includegraphics{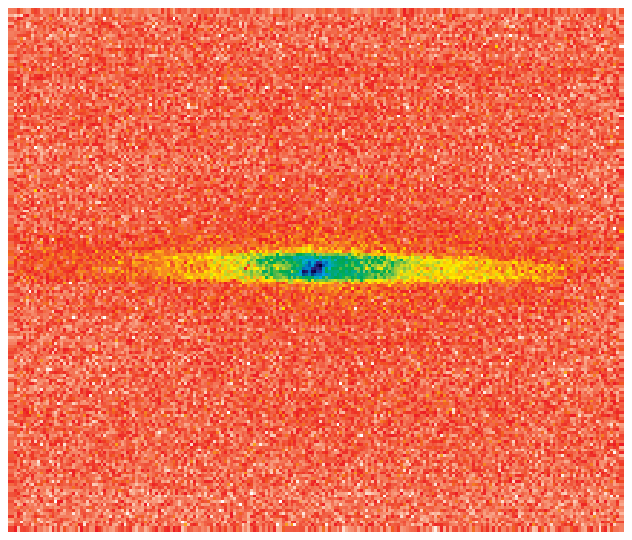}}
\caption{Head-on image of the channel created by a laser beam traversing the center of an optical dipole trapped atom sample.}
\label{Figure15}
\end{figure}

We begin with two laser beams, one near the 795 nm D1 line (the light shift (LS) laser) and one near the 780 nm D2 line (the probe). The beams are nearly spatially mode matched and propagate along the same axis.  They are focused so that their beam waist is approximately $15\; \mu$m with a Rayleigh range of approximately  $900\; \mu$m. The LS laser is used to create a channel using the AC Stark shift. In general, shifts in the ground and excited states must be considered so as to obtain the effective differential atomic resonance shift between the states. However, in performing calculations motivated by Safranova et al\cite{Safranova} and Griffin\cite{Griffin}, it was found that to a good approximation only the ground state is shifted around 795 nm. The light-shift can then be estimated as

\begin{equation}
\Delta E=\frac{\alpha_0P}{\pi\epsilon_0c{w_0}^2}
\end{equation}

where $\alpha_0$ is the scalar polarizability, $P$ is the power of the beam, $\epsilon_0$ is the permitivitty of free space, $c$ the speed of light, and $w_0$ the beam waist. This expression is written in SI units; division by Planck's Constant $h$ gives the shift in conventional frequency units.

The AC Stark shift can be measured by probing the spectral shift of hyperfine components of the D2 line. By changing the detuning or intensity of the LS laser, the ground state energy is changed, and this effect is seen by measuring the shift of resonance of the $F = 2 \rightarrow F' = 3$ transition of the D2 hyperfine multiplet. As seen in Fig. 16, detuning the LS laser 10 GHz below resonance with 5 mW of power leads to a shift of $\sim$20 MHz. Increasing the detuning to 100 GHz reduces the shift by nearly a factor of 10 to 2.4 MHz.

\begin{figure}
{\includegraphics{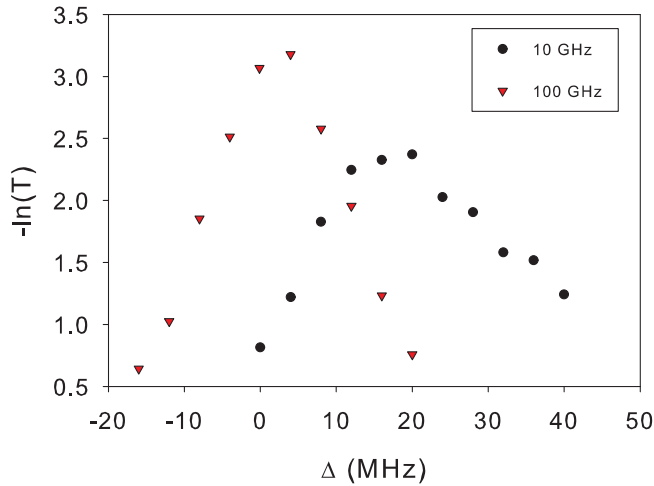}}
\caption{Spectral shifts and Zeeman broadening of atomic resonance due to the ac Stark shift generated by the light shift laser.  Data for two different detunings and a light shift laser power of about 5 mW is shown.}
\label{Figure16}
\end{figure}

Atoms within the LS beam path experience a resonance shift and the optical depth within this channel for light at the bare resonance for $5^2S_{1/2} \rightarrow 5^2P_{3/2}$ is reduced. The atoms outside of this channel create a dielectric wall which is optically deep and difficult for photons to penetrate. Thus, scattering is reduced in the transverse direction and mostly limited to forwards or backwards scattering, creating a quasi-1D system.

\subsubsection{Conclusion}
In this section, we have given an overview of one method by which to create in the laboratory a quasi one dimensional configuration for light propagation through a high density and cold atomic gas. We have demonstrated and partially characterized such a configuration for an optical dipole trapped gas of $^{87}Rb$ atoms. The overall scientific aim is to generate a soft cavity type arrangement for study of random lasing and Anderson light localization in reduced spatial geometry.

\section*{Acknowledgments}
We also appreciate financial support by the National Science Foundation (Grant Nos. NSF-PHY-0654226 and NSF-PHY-1068159), the Russian Foundation for Basic Research (Grant No. RFBR-CNRS 12-02-91056). D.V.K. would like to acknowledge support from the External Fellowship Program of the Russian Quantum Center (Ref. Number 86). We also acknowledge the generous support of the Ministry of Education and Science of the Russian Federation (State Assignment 3.1446.2014K).

\end{document}